\documentclass{ws-procs9x6}

\begin{document}

\title{PHASE STRUCTURE OF THERMAL QED 
BASED ON THE HARD THERMAL LOOP IMPROVED LADDER 
DYSON-SCHWINGER EQUATION \\
--A ``GAUGE INVARIANT'' SOLUTION--}

\author{HISAO NAKKAGAWA, HIROSHI YOKOTA and KOJI YOSHIDA}

\address{Institute for Natural Sciences, Nara University,\\
Nara 631-8502, Japan\\
E-mail: \{nakk, yokotah, yoshidak\}@daibutsu.nara-u.ac.jp}

\begin{abstract}
Based on the hard-thermal-loop resummed improved ladder Dyson-Schwinger equation for the
fermion mass function, we study how we can get the gauge invariant solution in the sense it satisfies
the Ward identity. Properties of the ``gauge-invariant'' solutions are discussed.
\end{abstract}


\bodymatter

\section{Introduction and summary}\label{aba:sec1}
The Dyson-Schwinger equation (DSE) is a powerful tool to investigate with the analytic procedure the
nonperturbative structure of field theories, such as the phase structure of gauge theories.
However, the full DSEs are coupled integral equtions for several unknown functions, thus are hard to be
solved without introducing appropriate approximations. We usually adopt the step-by-step approach to
this problem, firstly approximate the integration kernel by the tree, or, ladder kernel, next use the improved
ladder one, etc. Advantage of the DSE analysis lies in the possibility of such an systematic improvement through
the analytic investigation. 

Analyses of the DSE have proven to be successful in studying the phase structure of vacuum gauge theories [1-3].
In the Landau gauge DSE with the ladder kernel for the fermion mass function in the vacuum QED, the fermion
wave function renormalization constant is guaranteed to be unity [1], satisfying the Ward identity. Thus irrespective
of the problem of the ladder approximation, the results obtained would be gauge invariant.

Same analyses have been carried out in the finite temperature/density case with the ladder kernel [4-6], and with
the hard-thermal-loop (HTL) resummed improved ladder kernel [7].  Results of Ref. [7] show that at finite
temperature/density it is important to correctly analyse the physical mass function $\Sigma_R$, the mass function of the
 "unstable" quasiparticle in thermal field theories, and also to correctly take the dominant thermal effect into the
interaction kernel.   

All the preceding analyses [4-7], however, suffer from the serious problem coming from the ladder approximation of
the interaction kernel. Although in the vacuum case, despite the use of ladder kernel, in the analysis in the Landau
gauge the Ward identity is guaranteed to be satisfied, at finite temperature /density there is no such guarantee. In fact,
even in the Landau gauge the fermion wave function renormalization constant largely deviates from unity [7,8], being
not even real. At finite temperature/density the results obtained from the ladder DSE explicitly violate the Ward identity,
thus depend on the gauge, their physical meaning being obscure.

In this paper, we worked out, in the analysis of the HTL resummed improved ladder DS equation for the fermion mass function
in thermal QED, the procedure to get the gauge invariant solution in the sense it satisfies the Ward
identity, and investigated the properties of the "gauge invariant" solution. 

 Results of the present analysis are summarized as follows:

    (1) We can determine the solution that satisfies the Ward identity, namely the fermion wave function renormalization
constant being almost equal to unity. To get such a solution it is essential that the gauge parameter $\xi$ depends on the
momentum of the gauge boson.

    (2) The chiral phase transition in the massless thermal QED is confirmed to occur through the second order transition.

    (3) Two critical exponents $\nu$ and $\eta$ are consistent with constant within the range of values of temperatures
 and couplingconstants under analysis: $\nu = 0.395, \eta = 0.518$.

    (4) The effect of thermal fluctuation on the chiral symmetry breaking and/or restoration is smaller than that
expected in the previous analysis in the Landau gauge [7].

\section{The Dyson-Schwinger equation for the fermion self-energy function $\Sigma_R$}

The fermion self-energy function $\Sigma_R$ appearing in the fermion propagator 
\begin{equation}
 S_R(P)  = [ P\!\!\!\!/ + i \epsilon \gamma^0 - \Sigma_R(P)]^{-1} 
\end{equation}
can be decomposed at finite temperature and/or density as 
\begin{equation}
    \Sigma_R(P) =  (1 - A(P)) p_i \gamma^i - B(P) \gamma^0   + C(P)
\end{equation}
with $A(P)$, $B(P)$ and $C(P)$ being the three scalar invariants to be determined.  In the present analysis,
we use the HTL resummed
form $^*G^{\mu \nu}$ for the gauge boson propagator $G^{\mu \nu}$  
\begin{equation}
{}^*G^{\mu\nu} (K) = \frac{1}{{}^*\Pi_T -K^2 - i \epsilon k_0} A^{\mu \nu}
    + \frac{1}{{}^*\Pi_L -K^2 - i \epsilon k_0} B^{\mu \nu}
    - \frac{\xi}{K^2 + i \epsilon k_0} D^{\mu \nu} 
\end{equation}
where $^*\Pi_{L/T}$  is the HTL resummed longitudinal/transverse photon self-energy function [9], and
$A^{\mu \nu}$, $B^{\mu \nu}$ and $D^{\mu \nu}$ are the projection tensors [10], 
\begin{equation}
 A^{\mu \nu}=g^{\mu \nu} - B^{\mu \nu}- D^{\mu \nu},
 B^{\mu \nu}=- \tilde{K}^{\mu} \tilde{K}^{\nu}/K^2, 
 D^{\mu \nu}= K^{\mu} K^{\nu}/K^2, 
\end{equation}
where  $\tilde{K}=(k, k_0{\bf \hat{k}})$, $k=\sqrt{{\bf k}^2}$ 
and ${\bf \hat{k}}={\bf k}/k$ denotes the unit three vector along  ${\bf k}$.
 The parameter $\xi$ appearing in the term proportional to the projection tensor $D_{\mu \nu}$ represents the
gauge-fixing parameter 
($\xi=0$ in the Landau gauge). This gauge term plays an important role in the present analysis.

    The vertex function is approximated by the tree (point) vertex. With the instantaneous exchange approximation for the 
longitudinal photon mode, we get the DSEs for the three invariant functions $A(P)$, $B(P)$ and $C(P)$ 
\begin{eqnarray}
&  & -p^2[1-A(P)] = -e^2 \left. \int \frac{d^4K}{(2 \pi)^4}
       \right[ \{1+2n_B(p_0-k_0) \} \mbox{Im}[\ ^*G^{\rho \sigma}_R(P-K)]
       \times  \nonumber \\
  & & \Bigl[ \{ K_{\sigma}P_{\rho} + K_{\rho} P_{\sigma}
       - p_0 (K_{\sigma} g_{\rho 0} + K_{\rho} g_{\sigma 0} ) 
       - k_0 (P_{\sigma} g_{\rho 0} + P_{\rho} g_{\sigma 0} )
       + pkz g_{\sigma \rho} \nonumber \\
  & & + 2p_0k_0g_{\sigma 0}g_{\rho 0} \}\frac{A(K)}{[k_0+B(K)+i
       \epsilon]^2 - A(K)^2k^2 -C(K)^2 }
       + \{ P_{\sigma} g_{\rho 0} \nonumber \\
  & &  + P_{\rho} g_{\sigma 0} - 2p_0 g_{\sigma 0} g_{\rho 0} \}
       \frac{k_0+B(K)}{[k_0+B(K)+i \epsilon]^2 - A(K)^2k^2
       -C(K)^2 } \Bigr] 
       \nonumber \\ 
  & & + \{1-2n_F(k_0) \} \ ^*G^{\rho \sigma}_R(P-K) \mbox{Im} \Bigl[
       \{ K_{\sigma}P_{\rho}  + K_{\rho} P_{\sigma} - p_0 (K_{\sigma}
       g_{\rho 0} + K_{\rho} g_{\sigma 0} ) \nonumber \\
  & &  - k_0 (P_{\sigma}g_{\rho 0} + P_{\rho} g_{\sigma 0} ) + pkz g_{\sigma \rho} + 2p_0k_0g_{\sigma 0}g_{\rho 0}\}
       \times \nonumber \\
  & &  \frac{A(K)}{[k_0+B(K)+i \epsilon]^2 - A(K)^2k^2-C(K)^2 } 
       +  \{ P_{\sigma} g_{\rho 0} + P_{\rho} g_{\sigma 0} \nonumber \\
  & & \left. - 2p_0 g_{\sigma 0} g_{\rho 0} \}
       \frac{k_0+B(K)}{[k_0+B(K)+i \epsilon]^2 - A(K)^2k^2
       -C(K)^2 } \Bigr] \right] \ ,
\end{eqnarray}
\begin{eqnarray}
& & - B(P)= -e^2 \left. \int \frac{d^4K}{(2 \pi)^4} \right[
        \{1+2_B(p_0-k_0)\} \mbox{Im}[\ ^*G^{\rho \sigma}_R(P-K)] \times
         \nonumber \\
  & & \Bigl[ \{ K_{\sigma} g_{\rho 0} + K_{\rho} g_{\sigma 0}
       - 2k_0 g_{\sigma 0} g_{\rho 0} \}
       \frac{A(K)}{[k_0+B(K)+i \epsilon]^2 - A(K)^2k^2
       -C(K)^2 } \nonumber \\
  & & + \{ 2g_{\rho 0} 2g_{\sigma 0} - g_{\sigma \rho} \} 
       \frac{k_0+B(K)}{[k_0+B(K)+i \epsilon]^2 - A(K)^2k^2-C(K)^2 }
       \Bigr] \nonumber \\ 
  & & + \{1-2n_F(k_0) \} \ ^*G^{\rho \sigma}_R(P-K) \mbox{Im} \Bigl[ \frac{A(K)}{[k_0+B(K)+i
       \epsilon]^2 - A(K)^2k^2 -C(K)^2 } \nonumber \\
  & & \times \{ K_{\sigma} g_{\rho 0} + K_{\rho} g_{\sigma 0} - 2k_0 g_{\sigma 0} g_{\rho 0} \} \nonumber \\
  & &  \left. + \frac{k_0+B(K)}{[k_0+B(K)+
       i \epsilon]^2 - A(K)^2k^2-C(K)^2 }
       \{ 2g_{\rho 0} 2g_{\sigma 0} - g_{\sigma \rho} \} \Bigr] 
       \right] \ , \\
& &  C(P) = -e^2 \int \frac{d^4K}{(2 \pi)^4} g_{\sigma \rho} 
       \{1+2_B(p_0-k_0) \} \mbox{Im}[\ ^*G^{\rho \sigma}_R(P-K)]
       \times \nonumber \\
  & & \Bigl[ \frac{C(K)}{[k_0+B(K)+i \epsilon]^2 - A(K)^2k^2
       -C(K)^2 } + \{1-2n_F(k_0) \} \times \nonumber \\ 
  & & \left. \ ^*G^{\rho \sigma}_R(P-K) \mbox{Im} \Bigl[
       \frac{C(K)}{[k_0+B(K)+i \epsilon]^2 - A(K)^2k^2
       -C(K)^2 } \Bigr] \right] \ .
\end{eqnarray}

    The function $A(P)$ is nothing but the inverse of the fermion wave function renormalization constant $Z_2$,
thus must be unity in order to satisfy the Ward identity in the ladder DSE analysis, where the vertex function
receives no renormalization effect,  $Z_1 =1$.
 
     We must solve the above DSEs and get the solution satisfying the Ward identity $Z_2 = Z_1 (=1)$,
 where $Z_2 = A(P)^{-1}$. The procedure to get the ``gauge invariant'' solution is as follows;

     (1) Assume the nonlinear gauge such that the gauge parameter $\xi$ to be a function of the photon momentum
$K = (k_0, {\bf k})$, and parametrize $\xi$ as
\begin{equation}
         \xi(k_0, k) = \sum \xi_{mn} H_m(k_0) L_n(k) ,
\end{equation}
where $\xi_{mn}$  are unknown parameters to be determined, $H_m$ the Hermite functions and  $L_n$ the Laguerre functions.

    (2) In solving DSEs iteratively, impose the condition $A(P) =1$ by constraint for the input-functions at
each step of the iteration.

    (3) Determine $\xi_{mn}$  so as to minimize $|A(P) -1|^2$ for the out-put functions and find the solutions for
$B(P)$ and $C(P)$.
%
   
\section{Results of the analysis "Ÿ``gauge invariant'' solution"Ÿ}

Here we present the results obtained by allowing the gauge parameter $\xi$ to be a complex value. Number of parameters
$\xi_{mn}$ to minimize $|A(P) -1|^2$ is $2 \times 5 \times 2=20$ (i.e., $m=1 \sim 4$ and $n=1, 2$). All the quantities
with the mass dimension are evaluated in the unit of $\Lambda$, the cut-off parameter introduced as usual to regularize
the DSEs.


    Now we present the solution consistent with the Ward identity, i.e., the ``auge invariant'' solution. 
Firstly in Fig.1 we show $\mbox{Re}{A(P)}$.
For comparison, results in the constant $\xi$ analyses are also shown in the same figure. 
\begin{figure}
\begin{center}
\psfig{file=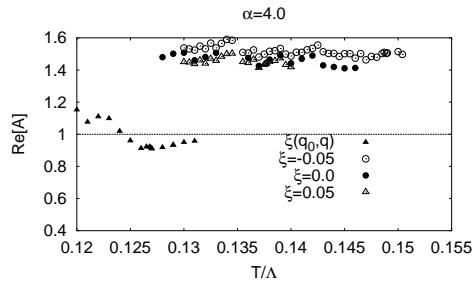,width=2.5in}
\end{center}
\caption{Comparison of the renormalization constant Re[$A$] at the coupling constant $\alpha=4.0$ evaluated at
$p_0=0$, $p=0.1$, see, text. }
\label{aba:fig2}
\end{figure}
                           
Next let us study the property of the phase transition. Fig.2(a) shows the real part of the fermion
mass $\mbox{Re}{M (P)}$, $M(P) \equiv  C(P)/A(P)$,
obtained from the ``gauge invariant'' solution, as a function of the temperature $T$ . The mass is
evaluated at $p_0=0$, $p=0.1$, to be consistent with
the standard prescription to define the mass in the static limit, $p_0=0$, $p \to 0$.
\begin{figure}
\begin{center}
\psfig{file=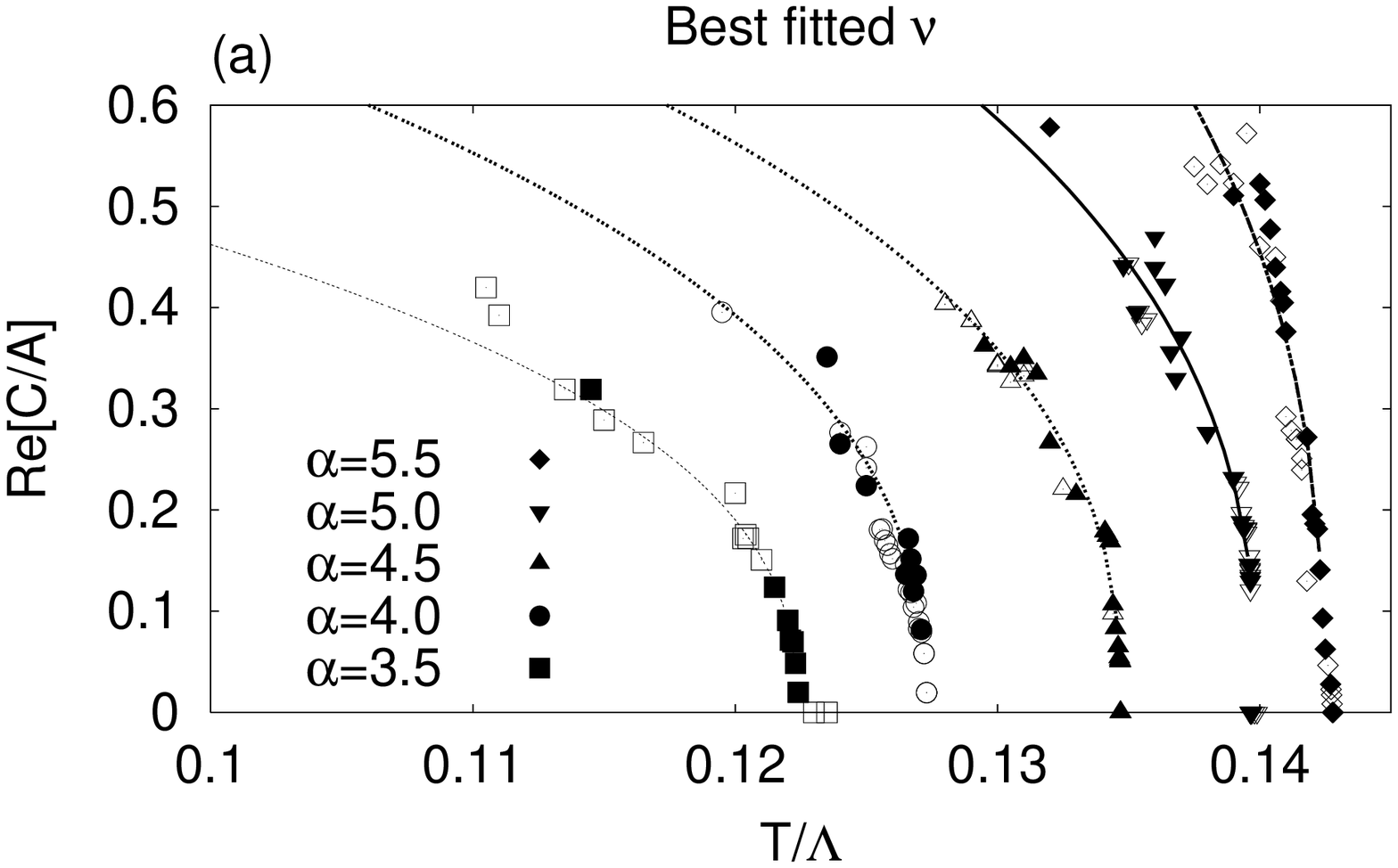,width=2.1in} \ \ \psfig{file=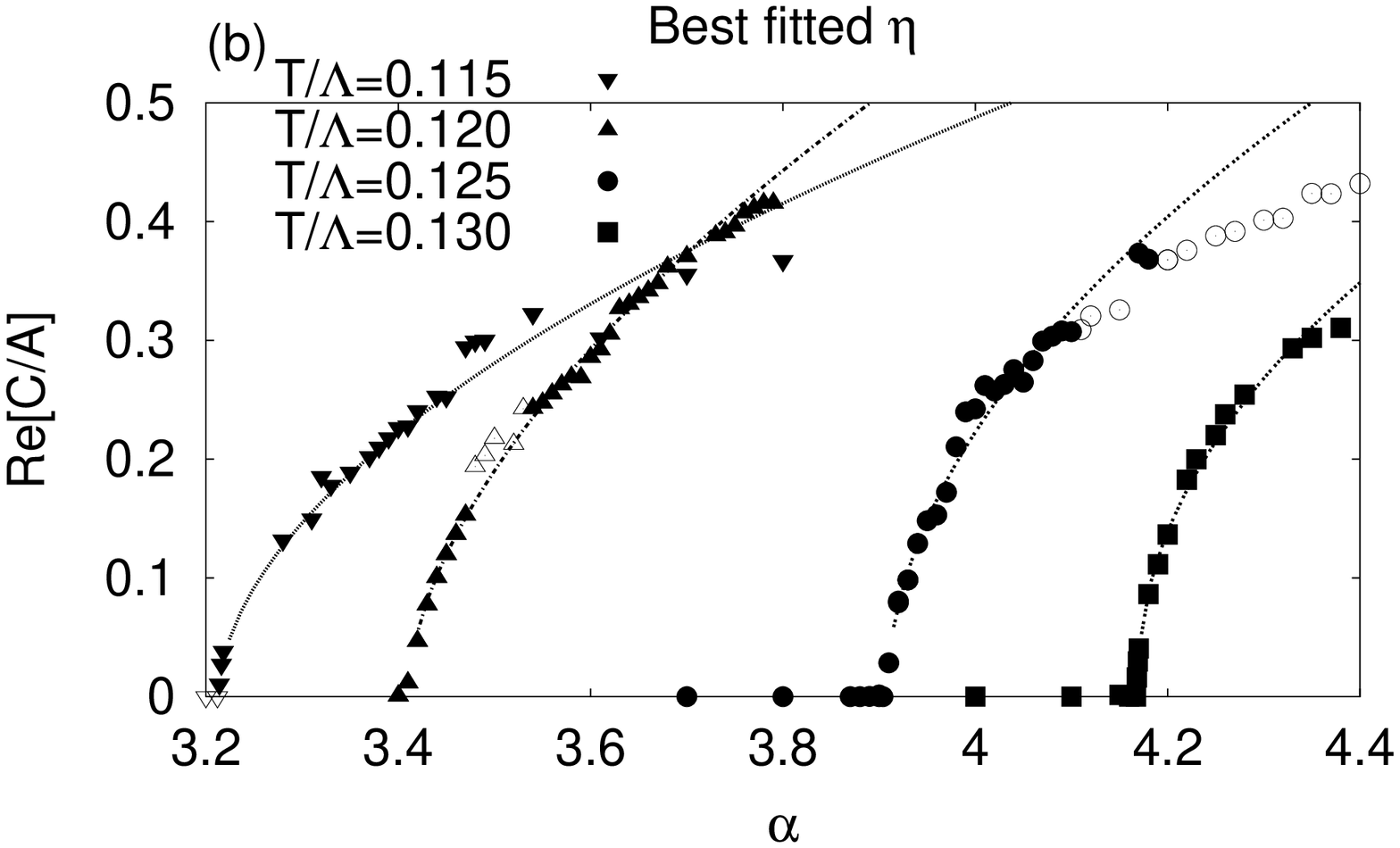,width=2.1in}
\end{center}
\caption{(a)Temperature-dependence of the fermion mass Re[$M(P)$] for various values of the coupling $\alpha$.
(b) Coupling constant-dependence of Re[$M(P)$] for various values
 of the temperature $T$. Both are evaluated at $p_0=0$ and $p=0.1$. As for the various curves, see text.}
\label{aba:fig3}
\end{figure}

 The curves in the figure show the best-fit curves in determining the critical temperature $T_c$ and the critical
exponent $\nu$, by fitting, at each coupling constant $\alpha$ and
near the critical temperature $T_c$ , the temperature-dependent data of $\mbox{Re}{M (P)}$ to the functional form
\begin{equation}
            \mbox{Re}{M (P)} = C_T(T_c - T)^{\nu} . 
\end{equation}

  Also shown in Fig.2(b) is the $\mbox{Re}{M (P)}$ obtained from the ``gauge invariant" solution, as a function of
the coupling constant $\alpha$. The mass is evaluated at $p_0=0$, $p=0.1$ as above. 
The curves in the figure show the best-fit curves in determining the critical coupling $\alpha_c$ and the critical
exponent $\eta$, by fitting, at each temperature T  and near the critical coupling $\alpha_c$,
the coupling-dependent data of $\mbox{Re}{M (P)}$ to the functional form 
\begin{equation}
            \mbox{Re}{M (P)} = C_{\alpha}(\alpha - \alpha_c)^{\eta} . 
\end{equation}

The determined critical exponents are given in Table 1.
\begin{table}[hbtp]
\tbl{Critical exponent $\nu$ for various values of the coupling constant $\alpha$ and
critical exponent $\eta$ for various values of the temperature $T$}
{\begin{tabular}{ccccc}
 $\alpha$ & $\nu$ & \ \ \ \ \ \ \ & $T$ & $\eta$ \\ \hline
   3.5 & 0.42800 &  & 0.115 & 0.54718 \\ 
   4.0 & 0.38126 &  & 0.120 & 0.57872 \\
   4.5 & 0.36420 &  & 0.125 & 0.51430 \\
   5.0 & 0.40579 &  & 0.130 & 0.46153 \\
\hline
\end{tabular}}
\end{table}

    The averaged value of $\nu$ over the various coupling is $<\mu> = 0.395$, 
which fits to all the data $\mbox{Re}{M (P)}$ in Fig.2(a) irrespective of the value of the coupling constant. 
The averaged value of $\eta$ over various temperatures is $<\eta> = 0.518$,
which fits to all the data $\mbox{Re}{M (P)}$ in Fig.2(b) irrespective of the value of the temperature.

     The phase boundary curve in the ($T,\alpha$)-plane thus determined shows that the region of the symmetry broken
phase shrinks to the low-temperature-strong- coupling side compared with that of the Landau gauge. This fact means
 that the effect of thermal fluctuation on the
 chiral symmetry breaking/restoration is smaller than that expected in the previous analysis in the Landau gauge [7]. 

\section{Discussion and comments}
Results presented in the present paper are preliminary, because of the rough analysis of the data processing.
 We are now refining the data analysis and 
soon get the results of the thorough reanalysis. Though the main conclusion will not be altered, several remarks
 should be added.
 
    (1) Present analysis was performed by allowing the gauge parameter $\xi$ to be a complex value. Such a choice
of gauge may correspond to studying the non-hermite dynamics, thus may cause some troubles. What happens if we
restrict the gauge parameter to the real value?  We are studying this case, finding a remarkable result:
In both cases results completely agree, thus getting a solution totally independent of the choice of gauges. 

     (2) In the present analysis, the consistency of the solution with the Ward identity is respected only by
imposing the condition $A(P) \approx 1$.  Needless to say, in solving the (improved) ladder Dyson-Schwinger
equation, there are no solutions totally consistent with the Ward identity. 
Despite that fact, following point should be closely examined: At least around or in the static limit, $p_0=0$,
$p \to 0$, where we calculated (defined) the mass, each invariant function $A(P)$, $B(P)$ or $C(P)$ should not
have big momentum dependence. This condition may be important in connection with the
consistency of the obtained solution with the  gauge invariance.  Result of the present analysis shows
that at least $B(P)$ and $C(P)$ satisfy this condition.

\end{document}